\documentclass[english,superscriptaddress,showkeys,onecolumn,notitlepage]{revtex4-1}
\usepackage[T1]{fontenc}
\usepackage[latin9]{inputenc}
\usepackage{color}
\usepackage{babel}
\usepackage{mathtools}
\usepackage{amsmath}
\usepackage{amssymb}
\usepackage{graphicx}
\usepackage[unicode=true,pdfusetitle,
 bookmarks=true,bookmarksnumbered=false,bookmarksopen=false,
 breaklinks=true,pdfborder={0 0 0},backref=false,colorlinks=true]
 {hyperref}
\hypersetup{
 citecolor=blue,linkcolor=cyan,urlcolor=blue,filecolor=green}
\usepackage{breakurl}

\usepackage{palatino}

\makeatletter
\@ifundefined{textcolor}{}
{
 \definecolor{BLACK}{gray}{0}
 \definecolor{WHITE}{gray}{1}
 \definecolor{RED}{rgb}{1,0,0}
 \definecolor{GREEN}{rgb}{0,1,0}
 \definecolor{BLUE}{rgb}{0,0,1}
 \definecolor{CYAN}{cmyk}{1,0,0,0}
 \definecolor{MAGENTA}{cmyk}{0,1,0,0}
 \definecolor{YELLOW}{cmyk}{0,0,1,0}
}
\usepackage{babel}

\makeatother

\begin{document}

\title{The sudden change phenomenon of quantum discord}

\author{Lucas C. C\'eleri}
\email{lucas@chibebe.org}
\address{Instituto de F\'{i}sica, Universidade Federal de Goi\'{a}s, 74.001-970, Goi\^{a}nia, Goi\'{a}s, Brazil}

\author{Jonas Maziero}
\email{jonas.maziero@ufsm.br}
\address{Departamento de F\'{i}sica, Centro de Ci\^{e}ncias Naturais e Exatas, Universidade Federal de Santa Maria, Avenida Roraima 1000, 97105-900, Santa Maria, RS, Brazil}
\address{Instituto de F\'{i}sica, Facultad de Ingenier\'{i}a, Universidad de la Rep\'{u}blica, J. Herrera y Reissig 565, 11300, Montevideo, Uruguay}

\begin{abstract}
Even if the parameters determining a system's state are varied smoothly, the behavior of quantum correlations alike to quantum discord, and of its classical counterparts, can be very peculiar, with the appearance of non-analyticities in its rate of change. Here we review this sudden change phenomenon (SCP) discussing some important points related to it: Its uncovering, interpretations, and experimental verifications, its use in the context of the emergence of the pointer basis in a quantum measurement process, its appearance and universality under Markovian and non-Markovian dynamics, its theoretical and experimental investigation in some other physical scenarios, and the related phenomenon of double sudden change of trace distance discord. Several open questions are identified, and we envisage that in answering them we will gain significant further insight about the relation between the SCP and the symmetry-geometric aspects of the quantum state space.
\end{abstract}

\keywords{quantum discord, open systems, decoherence, sudden change}

\maketitle
\tableofcontents{}

\section{Introduction}
\label{introduction}

In principle, it is always possible to find a local observable whose measurement does not disturb a system in a classical-incoherent state. However, the presence of quantum coherence in composite-correlated systems makes local interrogation without disturbance inconceivable \cite{Adesso_LQU}. Quantum discord (QD) is the generic name we give to correlations that have a quantum character. Actually, the minimum ``distance'' between the states of a multipartite system after and before a local measurement is quantified by discord-like functions \cite{RevKavan}. This kind of correlation is more general than quantum entanglement and has been shown to be important not only for the fundamentals of physics \cite{Lidar_CP,Wiseman,Liu_HSD_TPT,Wu_SR,You_MFs,Vennin_CMB} but also as a resource for quantum information science (QIS) \cite{Nielsen-Chuang,Preskill_LN,Wilde_book} applications. A few examples of QIS ``protocols'' for which there are strong evidences that QD is key for fueling the quantum advantage are noisy computation \cite{Datta_DQC1}, state merging \cite{Datta_DSM,winter_DSM}, assisted state discrimination \cite{Retamal_DSD}, cryptography \cite{Pirandola_QKD}, energy transport \cite{Lloyd_DET}, quantum illumination \cite{Vedral_DQI}, remote state preparation \cite{Vedral_DRSP}, data hidding \cite{Piani_DH}, and metrology \cite{Adesso_DMet,Fuentes_DMet}.

Although many useful theoretical developments are usually made in QIS by considering first idealized isolated systems, subsequently their realistic open dynamics must be taken into account \cite{Nielsen-Chuang,Preskill_LN,Wilde_book}. Quantum features such as entanglement and discord are particularly fragile to the influence of the environment. This kind of interaction usually smears the quantumness of the system of interest, inducing it to behave in a more classical manner. Such type of process is detrimental for applications of these quantities in QIS and generally goes under the name of decoherence. Motivated by that observation, much effort have been made in order to describe how useful quantum features are affected by several kinds of decoherent dynamics. In particular, the dynamic behavior of quantum discord has been shown to be very peculiar \cite{Maziero_scp}, with sudden discontinuities in its derivative being identified and analyzed in diverse physical scenarios.

This sudden change phenomenon (SCP) of classical and quantum correlations has been the theme of numerous works in the last few years, with considerable developments being put forward. Its geometric description and interpretation were addressed \cite{Caves_BDS,Fei_geo,Du_geo,Han_geo,Zhu_geo,Mateus_geo}. It has been regarded in the context of classical and quantum phase transitions \cite{Sarandy_pt_pra,Werlang_pt_prl,IBose_pt_jpb,Sarandy_qt_ijmpb,Campbell_pt_pra,Fanchini_pt_prb} and in the dynamics of critical systems \cite{Li_qt_dyn,Xu_qt_dyn,Qin_qt_dyn}. Considering Markovian \cite{XWang_mark,Fan_mark,Shi_mark,Long_mark,AWang_mark,Fei_mark,Ye_mark} and non-Markovian \cite{Non-Mark} decoherent dynamics, the establishment of system-environment correlations was investigated in Refs. \cite{An_SE,Han_SE,Li_SE,Zeng_SE,Zhang_SE}, while the protection of quantum correlations was adressed in Refs.  \cite{Xu_prot,Fanchini_prot,Maniscalco_prot,Xi_prot,Fan_prot}. Moreover, investigations of systems in relativistic motion \cite{Lucas_rel,Jing_rel,Ramzan_rel} or subjects to classical, chaotic, thermal, and other kinds of noise environments were also reported \cite{Feng_chao,Karpat_class,Liang_class,Li_ther,Kuang_ther,An_ther,Yue_ther,Guo_ther,Tian_comm,Liang_comm,Liao_comm,Huang_squee}. Studies considering interesting physical systems like quantum dots \cite{Berrada_qdots,Horodecki_qdots}, atoms interacting with separated or common cavities \cite{Xu_cav,Orszag_cav,Almeida_cav} or interacting via plasmonic waveguides \cite{Yao_plas,Paspalakis_plas} appeared in the literature. Finally, some interesting physical interpretations, as e.g. related to the emergence of the pointer basis in a quantum measurement process \cite{Paulo_epb}, or the worst-case-scenario fidelity of quantum teleportation \cite{Roszak_tel} and the quantum speed limits \cite{Xia_sl} shed new light on the subject. Here we shall make a survey of some important points related to these developments intending to open up the path for reaching a better physical and information-theoretic understanding of the SCP of classical and quantum correlations, bringing out thus its possible role in physics and in QIS.

\section{Quantum mechanics, open system dynamics and quantum correlations}

In this section, we present some important concepts of quantum mechanics and the quantification of correlations. We start, in Sec. \ref{QM}, by recalling the postulates of quantum mechanics, with particular emphasis on the description of open quantum dynamics via the Kraus' operator-sum representation. In Sec. \ref{zoo} we present the primary concepts involved in quantum, classical, and total correlation quantification and describe some of the correlation measures considered in this article. Our exposition shall be limited to discrete systems divided in two parts, but the main ideas can be straightforwardly translated to multipartite systems.

\subsection{Quantum mechanics of closed and open systems}
\label{QM}

In standard quantum mechanics \cite{Nielsen-Chuang}, the states of closed systems are described by normalized vectors on a projective Hilbert space $\mathcal{H}$. In Schr\"odinger's picture, these vectors evolve unitarily (and therefore reversibly): $|\psi_{t}\rangle=U|\psi\rangle$, with $U$ being a linear operator such that $UU^{\dagger}= U^{\dagger}U = \mathbb{I}$ and $\mathbb{I}|\psi\rangle=|\psi\rangle$ for all $|\psi\rangle\in\mathcal{H}$. The time evolution operator $U$ satisfies Schr\"odinger's equation: $i\hbar\partial_{t}U=HU$, where $H$ is the system Hamiltonian. Observables are described by Hermitian operators, $O=O^{\dagger}$. For a system prepared in state $|\psi\rangle$, measurements of $O$ at a time $t$ yield one of its real eigenvalues $o_{i}$ with probability given by Born's rule: $p_{i}=|\langle o_{i}|\psi_{t}\rangle|^{2}$ ($O = \sum_i o_i |o_{i}\rangle\langle o_{i}|$ is the spectral decomposition of $O$). Immediate subsequent measurements of $O$ always produce the same result. Because of this repeatability of measurements, one says that the system's state ``jumps'' to the observable's corresponding eigenvector $|o_{i}\rangle$, which will be the system's state immediately after the measurement. Thus, in quantum mechanics the measurement process will irreversibly change the state of the system unless it is prepared in one of the eigenstates of the observable being measured.

All systems in Nature generally interact with their surrounds, thus breaking local unitarity. Considering that quantum states lies at the core of quantum information science, this fact has been particularly important in the development of this theory, whose main focus is the manipulation of the information stored in the quantum degrees of freedom \cite{Nielsen-Chuang}. In this open system scenario, we may use the standard quantum description for the whole universe, but more general mathematical tools are needed when we focus on the description of some particular subsystem. In this case, because of the correlations generated between this system and the rest of the universe (the environment), the state of the first must be described by a convex mixture of state vectors, a density operator $\rho$, which is a positive-semidefinite, $\rho\ge 0$, linear operator with unit trace, $\mathrm{Tr}(\rho)=1$.

In what follows we briefly describe the dynamics of open quantum systems in terms of the so called operator sum representation \cite{Nielsen-Chuang}. Let $\mathcal{D}(\mathcal{H})$ be the space of density operators, $\mathcal{H}_{S}\otimes\mathcal{H}_{E}$ be the whole system Hilbert space, and $\rho=\rho^{S}\otimes|E_{0}\rangle\langle E_{0}|$ be the initial system-environment state. $\rho^{S}\in\mathcal{D}(\mathcal{H}_{S})$ is the system initial state and $\{|E_{i}\rangle\}$ is an orthonormal basis for $\mathcal{H}_{E}$. We can take the environment as being described by a pure state since we can always treat its purification without changing physical conclusions. 

Then, assuming that the entire system evolves by means of a unitary operation, $\rho_{t}=U\rho U^{\dagger}$, and considering the partial trace over the environment degrees of freedom we obtain 
\begin{equation}
\rho_{t}^{S}=\mathrm{Tr}_{E}(\rho_{t})={\textstyle \sum_{l}}K_{l}\rho^{S}K_{l}^{\dagger},
\label{eq:Kraus-OSR}
\end{equation}
which is the operator sum representation of the evolved system state \cite{Kraus_AP,Maziero_Kraus,Maziero_ptr}. The set $\{ K_{i}\}$ are the Kraus operators, whose matrix elements in the basis $|S_{i}\rangle\in\mathcal{H}_{S}$ are defined as 
\begin{equation}
\langle S_{m}|K_{l}|S_{n}\rangle\equiv\langle S_{m}E_{l}|U|S_{n}E_{0}\rangle.
\end{equation}
In the last equation and hereafter we shall use the notation $|\phi\psi\rangle\coloneqq|\phi\rangle\otimes|\psi\rangle$. One can easily verify that $\sum_{l}K_{l}^{\dagger}K_{l}=\mathbb{I}_{S}$, what implies that the evolution (\ref{eq:Kraus-OSR}) is trace-preserving, i.e. $\mathrm{Tr}(\rho_{t}^{S})=1$ for all times. Equation (\ref{eq:Kraus-OSR}) is the most general quantum evolution, known as a completely-positive and trace-preserving map.

\subsection{Quantum discord quantifiers}
\label{zoo}

Let's consider a bipartite system with Hilbert space $\mathcal{H}_{a}\otimes\mathcal{H}_{b}$. When can the correlations between the parties $a$ and $b$ be regarded as being classical? One way to answer this question is by recalling that coherent superpositions are a fundamental character of quantum systems \cite{Feynman3}; and that the measurement of an observable, for a system prepared in a state other than its eigenstates, will involve wave function collapse and disturbance. So, it is reasonable to say that if the state of a bipartite system is invariant, not disturbed, under the composition of (non-selective) local projective measurement maps (LPMMs), i.e., if $\exists \,\Pi_{a}, \Pi_{b}\mid\Pi_{a}\circ\Pi_{b}(\rho)=\rho$, then the correlations between its constituent parts are of a classical nature. In the above equations, a LPMM applied to sub-system $a$ is defined as 
\begin{equation}
\Pi_{a}(\rho):={\textstyle \sum_{j}}\Pi_{j}^{a}\otimes\mathbb{I}_{b}\rho\Pi_{j}^{a}\otimes\mathbb{I}_{b},
\end{equation}
with $\sum_{j=1}^{\dim\mathcal{H}_{a}}\Pi_{j}^{a}=\mathbb{I}_{a}$ and $\Pi_{j}^{a}\Pi_{k}^{a}=\delta_{jk}\Pi_{j}^{a}$; and the analogous definition follows for $\Pi_{b}(\rho)\coloneqq\sum_{j}\mathbb{I}_{a}\otimes\Pi_{j}^{b}\rho\mathbb{I}_{a}\otimes\Pi_{j}^{b}$.

One can easily verify that, for the so dubbed classical-classical states,
\begin{equation}
\rho_{cc}={\textstyle \sum_{j,k}}p_{jk}\Pi_{j}^{a}\otimes\Pi_{k}^{b},\label{eq:rhoCC}
\end{equation}
where $p_{jk}$ is an arbitrary probability distribution ($p_{jk}\ge0$ and $\sum_{j,k}p_{jk}=1$), we have $\Pi_{a}\circ\Pi_{b}(\rho_{cc})=\rho_{cc}$. For a system prepared in such a state, there is no quantum uncertainty associated with measurements of the local observables $A=\sum_{j}a_{j}\Pi_{j}^{a}$ and $B=\sum_{k}b_{k}\Pi_{k}^{b}$. Following these lines, the classical-quantum states (invariance under a LPMM on $a$: $\Pi_{a}(\rho_{cq})=\rho_{cq}$),
\begin{equation}
\rho_{cq}={\textstyle \sum_{j}}p_{j}\Pi_{j}^{a}\otimes\rho_{j}^{b},\label{eq:rhoCQ}
\end{equation}
and the quantum-classical states (not disturbed by a LPMM on
$b$: $\Pi_{b}(\rho_{qc})=\rho_{qc}$), 
\begin{equation}
\rho_{qc}={\textstyle \sum_{j}}p_{j}\rho_{j}^{a}\otimes\Pi_{j}^{b},\label{eq:rhoQC}
\end{equation}
are also important for the theory and applications of quantum discord quantifiers. In the last two equations, $p_{j}$ is a probability distribution and $\rho_{j}^{s}\in\mathcal{D}(\mathcal{H}_{s})$ are density operators for $s=a,b$. We observe that in addition to the nondisturbability-based characterization above, the classicality of the correlations between the subsystems $a$ and $b$, prepared in either one of these three classes of states, can be given an operational interpretation in terms of the possibility of locally broadcast them \cite{Horodecki_NLB,Luo_NLB}.

Now that we have defined what may be considered as being the classical states, i.e., those states not possessing quantum correlations, in order to quantify the amount of quantumness in the correlations of a generic bipartite density operator $\rho$, we can use the distance or distinguishability between $\rho$ and the regarded classical states. But using any classical state would be ambiguous; thus we utilize the classical states which ``seems more'' like $\rho$. That is to say, we define 
\begin{eqnarray}
 &  & D_{d}(\rho)\coloneqq\min_{\rho_{cc}}d(\rho,\rho_{cc})\equiv\min_{p_{jk},\Pi_{j}^{a},\Pi_{j}^{b}}d(\rho,{\textstyle \sum_{j,k}}p_{jk}\Pi_{j}^{a}\otimes\Pi_{k}^{b}),\label{eq:DD}\\
 &  & D_{d}^{a}(\rho)\coloneqq\min_{\rho_{cq}}d(\rho,\rho_{cq})\equiv\min_{p_{j},\Pi_{j}^{a},\rho_{j}^{b}}d(\rho,{\textstyle \sum_{j}}p_{j}\Pi_{j}^{a}\otimes\rho_{j}^{b}),\label{eq:DDa}\\
 &  & D_{d}^{b}(\rho)\coloneqq\min_{\rho_{qc}}d(\rho,\rho_{qc})\equiv\min_{p_{j},\rho_{j}^{a},\Pi_{j}^{b}}d(\rho,{\textstyle \sum_{j}}p_{j}\rho_{j}^{a}\otimes\Pi_{j}^{b}),\label{eq:DDb}
\end{eqnarray}
where $d$ is a distance or distinguishability measure for density operators.

It is worthwhile noticing that there is a related, but somewhat more operational way to define discord quantifiers. In this approach one starts by recognizing that a LPMM transforms $\rho$ into a classical state. Thus, QD is defined as the distance between a state $\rho$ and the closest classical state obtained by applying a LPMM to it, i.e., 
\begin{eqnarray}
 &  & \mathcal{D}_{d}(\rho)=\min_{\Pi_{a},\Pi_{b}}d(\rho,\Pi_{a}\circ\Pi_{b}(\rho)),\label{eq:D}\\
 &  & \mathcal{D}_{d}^{a}(\rho)=\min_{\Pi_{a}}d(\rho,\Pi_{a}(\rho)),\label{eq:Da}\\
 &  & \mathcal{D}_{d}^{b}(\rho)=\min_{\Pi_{b}}d(\rho,\Pi_{b}(\rho)).\label{eq:Db}
\end{eqnarray}
These quantifiers capture the essence of quantum discord, which is equal to the minimum amount of correlation that is inevitably erased by a LPMM. It is noticeable that there is no extremization over probability distributions or over local states in these last three expressions. This is so because, in this case, they are determined by $\rho$ and by the LPMM. It is interesting observing that we would have $\mathcal{D}_{d}\equiv D_{d}$ if we set $p_{jk}=\mathrm{Tr}(\Pi_{j}^{a}\otimes\Pi_{k}^{b}\rho)$ in the equation for $D_{d}$. Besides, $\Pi_{a}(\rho)\equiv\rho_{cq}$ if we make $p_{j}=\mathrm{Tr}(\Pi_{j}^{a}\otimes\mathbb{I}_{b}\rho)$ and $\rho_{j}^{b}=\mathrm{Tr}_{a}(p_{j}^{-1}\Pi_{j}^{a}\otimes\mathbb{I}_{b}\rho\Pi_{j}^{a}\otimes\mathbb{I}_{b})$ in Eq. (\ref{eq:DDa}) and $\Pi_{b}(\rho)\equiv\rho_{qc}$ if we use $p_{j}=\mathrm{Tr}(\mathbb{I}_{a}\otimes\Pi_{j}^{b}\rho)$ and $\rho_{j}^{a}=\mathrm{Tr}_{b}(p_{j}^{-1}\mathbb{I}_{a}\otimes\Pi_{j}^{b}\rho\mathbb{I}_{a}\otimes\Pi_{j}^{b})$ in Eq. (\ref{eq:DDb}). But now let us address an important issue about these two types of QD quantifiers, one of then that, to the best of our knowledge, has not been addressed in literature. It is true, for instance, that any state in the class $\rho_{cc}$ can in principle be produced. Nevertheless, it is also well known that the probability distribution induced by local measurements (PDILM) on a system prepared in a certain state $\rho$ cannot, in general, simulate all PDILM. One famous example of this fact is found in the Bell nonlocality scenario \cite{Wehner_RMP}. Therefore, in addition to make the optimization problem more involved, the measures in Eqs. (\ref{eq:DD}), (\ref{eq:DDa}), and (\ref{eq:DDb}) may lead to misleading results due to the application of classical states possible ``unrelated'' to $\rho$, i.e., that cannot be obtained from $\rho$ via LPMMs.

After a LPMM is applied, $a$ and $b$ can still be correlated if, e.g., $p_{jk}\ne p_{j}^{a}p_{k}^{b}$, with $p_{j}^{a}$ and $p_{k}^{b}$ being probability distributions. This remaining correlation may be said to have a classical nature. But then, before the LPMM, the system posses two kinds of correlation. And it would be nothing but natural assuming that both types of correlation, the classical correlation (CC) and the quantum discord, add up to give the total correlation (TC) in $\rho$. Surprisingly, the panorama of the theory for measures of CCs and of TCs is even less satisfactorily developed than for QD. Next we regard some possible approaches that may be followed for the quantification of TC and of CC in bipartite states. A good starting point for that is the definition of uncorrelated states. If the sub-systems $a$ and $b$ are prepared independently, respectively, in states $\varpi_{a}$ and $\varpi_{b}$, then their joint density operator would be the product state $\varpi_{a}\otimes\varpi_{b}$. Similarly to what was done for QD, these states can be used to define a measure of TC
\begin{equation}
I_{d}(\rho)\coloneqq\min_{\varpi_{a},\varpi_{b}}d(\rho,\varpi_{a}\otimes\varpi_{b}).\label{eq:IId}
\end{equation}
We use $I_{d}$ instead of $T_{d}$ because of the historic and present importance mutual information (MI) has for TC quantification. Actually, the MI, which is defined as 
\begin{equation}
I(\rho)\coloneqq S(\rho_{a})+S(\rho_{b})-S(\rho)\label{eq:MI}
\end{equation}
with 
\begin{equation}
S(x)\coloneqq-\mathrm{Tr}(x\log_{2}x)\label{eq:vNE}
\end{equation}
being the von Neumann's entropy, is the only TC measure already given an operational meaning \cite{Winter_TC}.

It is intuitively expected that the closest product state of $\rho$ is obtained from the tensor product of its reductions. And for the distinguishability measure named quantum relative entropy \cite{Vedral_RE},
\begin{equation}
d_{re}(\rho,\xi)\coloneqq\mathrm{Tr}(\rho(\log_{2}\rho-\log_{2}\xi)),\label{eq:qre}
\end{equation}
this is indeed the case \cite{Willianson_DRE,Maziero_MI}, i.e., 
\begin{equation}
I_{re}(\rho)\coloneqq\min_{\varpi_{a},\varpi_{b}}d_{re}(\rho,\varpi_{a}\otimes\varpi_{b})=d_{re}(\rho,\rho_{a}\otimes\rho_{b})\equiv I(\rho).
\end{equation}
So, this TC is equal to mutual information. It is rather curious that e.g. for the trace distance, 
\begin{equation}
d_{tr}(x,y)\coloneqq||x-y||_{tr},
\end{equation}
where 
\begin{equation}
||x||_{tr}\coloneqq\mathrm{Tr}\sqrt{x^{\dagger}x}
\end{equation}
is the trace norm of $x$, the state $\rho_{a}\otimes\rho_{b}$ is not in general the uncorrelated state most similar to $\rho$ \cite{Adesso_DTC_Tr}. This issue certainly deserves additional-thoroughly investigation. This fact also motivates using a more operational definition also for TC. One possibility would be defining 
\begin{equation}
\mathcal{I}_{d}(\rho)\coloneqq\min_{\Omega}d(\rho,\Omega(\rho)),\label{eq:Id}
\end{equation}
where 
\begin{equation}
\Omega(\rho)\coloneqq{\textstyle \sum_{j}}p_{j}(U_{j}^{a}\otimes U_{j}^{b})\rho(U_{j}^{a}\otimes U_{j}^{b})^{\dagger}
\end{equation}
is a randomizing (decoupling) map leading any $\rho$ into product states and $U_{j}^{s}$ are appropriately chosen unitary operators acting on $\mathcal{H}_{s}$. Although being more difficult to calculate, the total correlation in Eq. (\ref{eq:Id}) may be more suitable because in general (some property of) $\rho$ may restrict the possible product states that can be produced by $\Omega$ acting on it.

What about classical correlation quantifiers? One possibility is arguing for additivity for correlations and simply define
\begin{equation}
C_{d}(\rho)\coloneqq I_{d}(\rho)-D_{d}(\rho).
\end{equation}
Or one can use the classical state minimizing the equation for e.g. $D_{d}$, and define the CC of $\rho$ as the TC of this state. We could continue presenting several other ways in which CC may be defined. Actually, this observation is in its own an indication that these definitions are not, in general, operationally satisfying. The only CC quantifier with a well defined information theoretic interpretation that we know of was inspired by the Holevo bound and was proposed by Henderson and Vedral in Ref. \cite{Vedral_D_jpa}. But before presenting their CC measure, let us consider the following equivalent distinguishability-based quantifier
\begin{equation}
C_{hv}^{a}(\rho)\coloneqq\max_{\Pi_{a}}I_{re}(\Pi_{a}(\rho))\equiv\max_{\Pi_{a}}I(\Pi_{a}(\rho))=\max_{\Pi_{j}^{a}}[S({\textstyle \sum_{j}}p_{j}\Pi_{j}^{a})+S({\textstyle \sum_{j}}p_{j}\rho_{j}^{b})-S({\textstyle \sum_{j}}p_{j}\Pi_{j}^{a}\otimes\rho_{j}^{b})].
\end{equation}
As the states $\Pi_{j}^{a}\otimes\rho_{j}^{b}$ have support in orthogonal subspaces of $\mathcal{H}_{a}\otimes\mathcal{H}_{b}$, one can verify that 
\begin{equation}
S({\textstyle \sum_{j}}p_{j}\Pi_{j}^{a}\otimes\rho_{j}^{b})=H(p_{j})+{\textstyle \sum_{j}}p_{j}S(\Pi_{j}^{a}\otimes\rho_{j}^{b})=S({\textstyle \sum_{j}}p_{j}\Pi_{j}^{a})+{\textstyle \sum_{j}}p_{j}S(\rho_{j}^{b}),
\end{equation}
where $H(p_{j})\coloneqq-\sum_{j}p_{j}\log_{2}p_{j}$ is Shannon's entropy. With this we get 
\begin{equation}
C_{hv}^{a}(\rho)=\max_{\Pi_{j}^{a}}[S({\textstyle \sum_{j}}p_{j}\rho_{j}^{b})-{\textstyle \sum_{j}}p_{j}S(\rho_{j}^{b})],
\end{equation}
which is the Holevo quantity maximized over LPMMs on particle $a$, which are used here to acquire information about the subsystem $b$. Now, we recall that any \emph{local quantum operation} can be written as 
\begin{equation}
\$_{a}(\rho)={\textstyle \sum_{j}}K_{j}^{a}\otimes\mathbb{I}_{b}\rho K_{j}^{a\dagger}\otimes\mathbb{I}_{b},\label{eq:LQO}
\end{equation}
with $\sum_{j}K_{j}^{a\dagger}K_{j}^{a}=\mathbb{I}_{a}$. As expected, $\$_{a}$ does not change the state of $b$: 
\begin{eqnarray}
\mathrm{Tr}_{a}(\$_{a}(\rho)) & = & {\textstyle \sum_{l}}\langle a_{l}|\otimes\mathbb{I}_{b}{\textstyle \sum_{j}}K_{j}^{a}\otimes\mathbb{I}_{b}\rho K_{j}^{a\dagger}\otimes\mathbb{I}_{b}|a_{l}\rangle\otimes\mathbb{I}_{b}={\textstyle \sum_{l,j}}\langle a_{l}|K_{j}^{a}({\textstyle \sum_{m}}|a_{m}\rangle\langle a_{m}|)\otimes\mathbb{I}_{b}\rho K_{j}^{a\dagger}|a_{l}\rangle\otimes\mathbb{I}_{b}\nonumber \\
 & = & {\textstyle \sum_{l,j,m}}\langle a_{m}|\otimes\mathbb{I}_{b}\rho K_{j}^{a\dagger}|a_{l}\rangle\langle a_{l}|K_{j}^{a}|a_{m}\rangle\otimes\mathbb{I}_{b}={\textstyle \sum_{m}}\langle a_{m}|\otimes\mathbb{I}_{b}\rho|a_{m}\rangle\otimes\mathbb{I}_{b}=\mathrm{Tr}_{a}(\rho)=\rho_{b}.
\end{eqnarray}
As $\Pi_{a}$ is a particular kind of quantum operation, then $\rho_{b}=\mathrm{Tr}_{a}(\rho)=\mathrm{Tr}_{a}(\Pi_{a}(\rho))=\sum_{j}p_{j}\rho_{j}^{b}$ and \cite{Vedral_D_jpa} 
\begin{equation}
C_{hv}^{a}(\rho)=S(\rho_{b})-\min_{\Pi_{j}^{a}}{\textstyle \sum_{j}}p_{j}S(\rho_{j}^{b}).\label{eq:CC_HV}
\end{equation}
This expression leads to a nice entropic interpretation for this CC. The state of $b$ is the mixture $\sum_{j}p_{j}\rho_{j}^{b}$ before and after $\Pi_{a}$ is applied to $a$. However, because of the correlations between $a$ and $b$, the information we get when $a$ collapses to $\Pi_{j}^{a}$ forces $b$ to be in one of the states $\rho_{j}^{b}$. And, on average, this decreases our uncertainty about the state of $b$.

The notion of quantum discord as a quantum correlation quantifier was first introduced by Ollivier and Zurek in Ref. \cite{Zurek-D-PRL} and is directly related to the Henderson-Vedral CC: 
\begin{equation}
D_{oz}^{a}(\rho)\coloneqq I(\rho)-C_{hv}^{a}(\rho).\label{eq:DiscOZ}
\end{equation}
The motivation for the name ``quantum discord'' comes from the fact that classically the two definitions for correlation, measured by mutual information, are equivalent, but in the quantum realm we can have $I(\rho)\ne C_{hv}^{a}(\rho)$.

As we have seen in this sub-section, there are several motivations one can follow to define a quantum discord quantifier. Of course, the same holds for classical and total correlations. And even within a certain ``kind'' of QD quantifier, we can employ, for instance, several dissimilarity measures, which will imply in multiple QD functions. Actually, many functions involving $\rho$ and e.g. $\Pi_{a}(\rho)$ can be used to defined QD quantifiers. For some examples of such quantities, see Refs. \cite{Horodecki_Deficit,Luo_MID,Terno_D_Maxwell,Brukner_HSD,Sarandy_GQD,Shaji_DEntropic,Adesso_LQU,Sarandy_D1,Adesso_D_Neg,Orszag_D_Bures,Giovannetti_D_DS,
Illuminati_DResp,Pati_SQD,Azmi_AHSD,Huang_EasyQDQ,Angelo_DPR,Wilde_DRenyi,Li_D_Weak} and references therein. This scenario has led to a variety of QDQs and has motivated the discussion about which properties they should enjoy \cite{Modi_CriteriaQD,Sarandy_Ambiguities}. But, as this is not the focus of this work, any other QD quantifier, and the related issues, will be introduced as needed in our subsequent analysis of the sudden change phenomenon.

\section{Uncovering of the SCP and its experimental verification }
\label{discovery}

The sudden change phenomenon was discovered in 2009 by us, R. M. Serra and V. Vedral, and was first reported in Ref. \cite{Maziero_scp}. We considered the simple situation with two qubits prepared initially in a Bell-diagonal state and let them to interact with local-independent environments. For the sake of definiteness, we will consider the so called Pauli channels \cite{Preskill_LN,Wilde_book}. These are the phase-damping channel, whose Kraus operators are 
\begin{equation}
K_{0}^{pd}=\sqrt{1-p}\,\mathbb{I}\mbox{,   } K_{1}^{pd}=\sqrt{p}\,|0\rangle\langle0| \mbox{  and  }K_{2}^{pd}=\sqrt{p}\,|1\rangle\langle1|,
\label{eq:ko_pd}
\end{equation}
the bit flip, described by the operators  
\begin{equation}
K_{0}^{bf}=\sqrt{1-p}\,\mathbb{I}\mbox{  and  }K_{1}^{bf}=\sqrt{p}\,\sigma_{1},
\label{eq:ko_bf}
\end{equation}
phase flip, 
\begin{equation}
K_{0}^{pf}=\sqrt{1-p}\,\mathbb{I} \mbox{  and  } K_{1}^{pf}=\sqrt{p}\,\sigma_{3}
\label{eq:ko_pf}
\end{equation}
and, finally, the bit-phase flip
\begin{equation}
K_{0}^{bpf}=\sqrt{1-p}\,\mathbb{I} \mbox{  and  } K_{1}^{bpf}=\sqrt{p}\,\sigma_{2}.
\label{eq:ko_bpf}
\end{equation}
$p$ is the parametrized time, usually written as $p = 1 - e^{-\gamma t}$, with $\gamma$ being the relaxation rate associated with the environment. These quantum channels have the property that they preserve the Bell-diagonal form of the evolved state, i.e., 
\begin{equation}
\rho_{p}^{bd}={\textstyle \sum_{i,j}}K_{i}^{qc}\otimes K_{j}^{qc}\rho_{0}^{ab}K_{i}^{qc^{\dagger}}\otimes K_{j}^{qc^{\dagger}}=2^{-2}\left(\mathbb{I}^{ab}+\mathbf{\Xi}_{p}^{qc}\cdot\mathbf{\Upsilon}\right).\label{eq:BDS}
\end{equation}
In the last equation $\mathbf{\Upsilon}=(\Upsilon_{1},\Upsilon_{2},\Upsilon_{3})\coloneqq(\sigma_{1}\otimes\sigma_{1},\sigma_{2}\otimes\sigma_{2},\sigma_{3}\otimes\sigma_{3})$ and the evolved correlation vectors $\mathbf{\Xi}_{p}^{qc}=(c_{1}^{(p)},c_{2}^{(p)},c_{3}^{(p)})$, with $c_{j}^{(p)}=\mathrm{Tr}(\rho_{p}^{ab}\Upsilon_{j})$, are given by 
\begin{eqnarray}
\mathbf{\Xi}_{p}^{pd} & = & \left(c_{1}(1-p)^{2},c_{2}(1-p)^{2},c_{3}\right),\\
\mathbf{\Xi}_{p}^{bf} & = & \left(c_{1},c_{2}(1-2p)^{2},c_{3}(1-2p)^{2}\right),\\
\mathbf{\Xi}_{p}^{pf} & = & \left(c_{1}(1-2p)^{2},c_{2}(1-2p)^{2},c_{3}\right),\\
\mathbf{\Xi}_{p}^{bpf} & = & \left(c_{1}(1-2p)^{2},c_{2},c_{3}(1-2p)^{2}\right).
\end{eqnarray}

\begin{figure}
\begin{centering}
\includegraphics[scale=0.3]{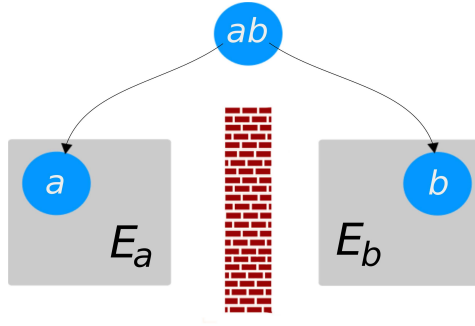}
\par\end{centering}
\caption{Illustration of the situation in which two systems are initially prepared
in a quantum-correlated state and afterwards let to interact with
local-independent environments.}
\label{aEabEb}
\end{figure}

For the Bell-diagonal class of states, the Ollivier-Zurek quantum discord (symmetric or asymmetric) can be computed analytically and reads \cite{Luo_DiscAn1,Luo_DiscAn2,Maziero_DiscAn}: 
\begin{equation}
D_{oz}(\rho_{p}^{bd})=I(\rho_{p}^{bd})-C_{hv}(\rho_{p}^{bd})={\textstyle \sum_{i,j=0}^{1}}\lambda_{ij}\log_{2}(4\lambda_{ij})-2^{-1}{\textstyle \sum_{i=0}^{1}}\left(1+(-1)^{i}c_{p}\right)\log_{2}\left(1+(-1)^{i}c_{p}\right),\label{eq:D_BDS}
\end{equation}
where $\lambda_{ij}=2^{-2}\left(1+(-1)^{i}c_{1}^{(p)}-(-1)^{i+j}c_{2}^{(p)}+(-1)^{j}c_{3}^{(p)}\right)$ and 
\begin{equation}
c_{p}=\max(|c_{1}^{(p)}|,|c_{2}^{(p)}|,|c_{3}^{(p)}|).\label{eq:c_scp}
\end{equation}

\begin{figure}[h]
\begin{centering}
\includegraphics[scale=0.5]{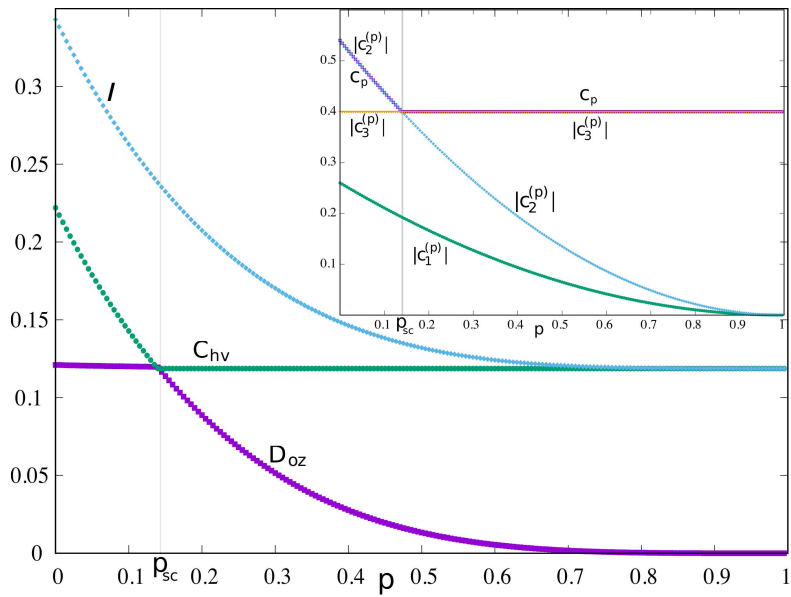} 
\par\end{centering}
\caption{Example of sudden change phenomenon taking place when an initial Bell-diagonal state with correlation vector $\mathbf{\Xi}_{0}=(-0.26,0.54,0.40)$ undergoes time evolution under local phase damping channels. Here, the sudden change time is $p_{sc}\approx0.14$. The inset shows the time evolution of the components of the correlation vector and $c_{p}$. We see that $|c_{3}^{(p)}|\lessgtr|c_{2}^{(p)}|$ for $p\lessgtr p_{sc}$. So, as the rate of change of these two components is different, so is the decaying rate of the classical, $C_{hv}$, and quantum, $D_{oz}$, correlation in the two quite distinct dynamical behavior regimes. For the environments and correlation measures regarded in this section, the classical correlation is constant from $p_{sc}$ thereon and there is only one possible value for $p_{sc}$, what shall be relevant in the discussion about the emergence of the pointer basis included in Sec. \ref{pointer}. Because the mutual information $I=C_{hv}+D_{oz}$ is a monotonically decaying function of $p$, for $p>p_{sc}$ the decaying rate of the quantum discord must be equal to that of the MI: $\partial_{p}D_{oz}=\partial_{p}I$. This fact indicates that if $\partial_{p}C_{hv}=\partial_{p}I$ for $p<p_{sc}$, then we shall have what has been called the freezing of quantum discord or the classical decoherence regime.}
\label{scp_pd} 
\end{figure}

For the sake of understanding the \emph{mathematical origin} of the SCP from these equations, we first call the attention for the fact that the components of $\mathbf{\Xi}_{p}^{qc}$ change differently with time, with their decaying rates depending on the kind of environment the system is interacting with. So, fixed an initial state $\mathbf{\Xi}_{0}^{qc}$ and the quantum channel, with exception of those $\mathbf{\Xi}_{0}^{qc}$ with the constant component null or greater than all the others, at the time dubbed sudden change time, $p_{sc}$, the index of the coefficient with greater modulus shall become different. And this can happens only if the change rate of $c_{p<p_{sc}}$ is different from that of $c_{p>p_{sc}}$ (see the example in Fig. \ref{scp_pd}). Well, and in this scenario it is this change of decay rate of $c_{p}$ that leads to the sudden change phenomenon of the classical correlation, $C_{hv}$, and of quantum discord, $D_{oz}$. Note that once the quantum mutual information, $I$, does not depend on $c_{p}$, it decays monotonically with time, without abrupt changes.

For all channels mentioned in this section, the SCP happens when $|\kappa|=\max(|\kappa'|,|\kappa''|)$, with $\kappa$ being the constant component of the correlation vector and $\kappa'$ and $\kappa''$ being the components of $\mathbf{\Xi}_{0}^{qc}$ that shall be affected by the interaction with the environment. Thus, in the case of the phase damping channel, the \emph{sudden change time} is 
\begin{equation}
p_{sc}=1-\sqrt{\frac{|\kappa|}{\max(|\kappa'|,|\kappa''|)}}=1-\sqrt{\frac{|c_{3}|}{\max(|c_{1}|,|c_{2}|)}},
\end{equation}
and for the other three channels $p_{sc}=2^{-1}(1-\sqrt{|\kappa|/\max(|\kappa'|,|\kappa''|)})$. It is worthwhile noticing also that if $|\kappa|\ne1$ and $\kappa\ne0$, there always exists a quantum channel and an initial state for which $p_{sc}\in(0,1)$; thus the SCP is seen to be universal.

Now, in order to comprehend the \emph{physical origin} of the SCP, we start by observing that a positive value of QD is obligatorily linked to a minimum amount of correlation between the subsystems that has to be destroyed by the non-selective measurement of local-compatible observables. Although, to our knowledge, there is no proof for general states, in the case of Bell-diagonal states we can verify that at the sudden change time the less disturbing observable is also altered. We anticipate that the resource theory of asymmetry \cite{Spekkens_rta} can shed more light on the interpretation and possible applications of the SCP in the general scenario. Determining formally the relationship between the SCP and the resource theory of asymmetry is an important open problem.

Soon after been reported in Ref. \cite{Maziero_scp}, the SCP was verified experimentally by Xu \emph{et al.}, as described in Ref. \cite{Guo_scp}. They used as qubits the polarization degrees of freedom of two photons generated via parametric down conversion. One of the qubits was then subject to a simulated-controlled dephasing environment. Moreover, the authors also verified the transition between the classical and quantum decoherence regimes, that have been theoretically noticed in Ref. \cite{Maniscalco_scp}. The dynamics shown in Fig. \ref{scp_pd} is one example of such a transition, where for $p<p_{sc}$ ($p>p_{sc}$) only classical (quantum) correlations are affected by the interaction with the environment. In 2011, we and our co-workers observed experimentally these dynamical behaviors for the real dephasing and dissipative environments encountered in liquid state nuclear magnetic resonance, where the two qubits were encoded in the $^{1}\mbox{H}$ and $^{13}\mbox{C}$ spin-1/2 nuclei \cite{Auccaise_scp}. A detailed description of related theory and experiments can be found in Refs. \cite{Diogo_Redfield,Diogo_ptrsa,Maziero_Dbjp}.

\section{The SCP and the emergence of the pointer basis}
\label{pointer}

In quantum mechanics, for any observable we can think of, coherent superpositions of its eigenstates are the rule, not the exception. And, as QM lists several of the known rules of Nature, an immediate question we are urged to ask is: Why it is so hard seeing or maintaining physical systems in states that are coherent with relation to certain observables? Decoherence \cite{Zeh_dec}, einselection \cite{Zurek_pb}, and quantum Darwinism \cite{Zurek_dar} are theories that give fine answers to this question. An important conceptual hint for these theories can be traced back to 1935, when Schr\"odinger exposed the idea that when two particles get entangled, they cannot be described ``by endowing each of them with a representative of its own'' \cite{Schrodinger}. Decoherence theory recognizes that in the system-environment dynamics, in Zeh's words, ``Any sufficiently effective interaction will induce correlations ... defining in general a large value of entropy'' of the system \cite{Zeh_dec}. This increase in disorder is generally associated with the diminishing of quantum coherence. By its turn, environment-induced superselection, or einselection \cite{Zurek_pb}, explains why some observables, the so called pointer observables, are the chosen ones for classical reality. It is the commutativity of the system-environment interaction Hamiltonian with the pointer observables that leads to the robustness of its eigenstates; and to the disappearance of its superpositions. Quantum Darwinism identifies the special role information has in this scenario. In addition to be able to ``survive'' under the influence of the environment, the pointer states are the ones that most efficiently spread its information by getting correlated with many parts of the surroundings. This leads to the classical feature of the classical world: redundancy, i.e., several observers can have non-disturbing access to the same information \cite{Zurek_dar}.

The next natural question to be asked is about how much time it takes for the quantum-to-classical transition to occur. That is to say, we want to understand when the pointer basis emerges. This question is usually addressed in the following scenario. We consider a quantum system S and obtain information about it via another quantum system M, the measurement apparatus. To do that, a pre-measurement evolution must take place to correlate S and M, reflecting the possible observable values (or states) of S into the evolved states of M \cite{von_Neumann}. But S and M are, in general, not isolate; and the interaction with the environment E correlates all the three entities, leading to increased mixing of the joint state of S and M. We recall here that the pointer basis is said to have emerged if, in principle, we can perform a non-selective von Neumann measurement in this basis without disturbing the system being measured. The issue now is if such measurement must refer to S or to M. Starting with the first choice, the figure of merit that have been usually used to quantify the time for the emergence of classicality is called the decoherence time, or decoherence half-life. This quantity may represent the time taken for S to lose its coherence with relation to the pointer basis \cite{Zurek_td,Wineland_td,Angelo_epb,Trauzettel_epb}. It is here that the \emph{SCP enters the scene}. As reported in Ref. \cite{Paulo_epb}, once in practice we look at M to obtain information about S, it is the disappearance of the coherence of M in the pointer basis that marks the transition to classicality. Such disappearance can be identified using $p_{sc}$, the time in which a sudden change in the classical correlation $C_{hv}^{M}$ between S and M occurs, if $C_{hv}^{M}$ remains constant for later times \cite{Paulo_epb,Sarandy_2sc_exp,Sarandy_epb}. It is worthwhile mentioning that for non-Markovian environments, a more complex and interesting situation with the possibility for more than one pointer observable and metastable pointer bases was reported in Ref. \cite{Wallentowitz_epb}.

\section{Dynamics of correlations}
\label{Markovian}

As described earlier, the SCP was predicted by studying the non-unitary evolution of the system. In other words, considering that the system of interest interacts with uncontrolled degrees of freedom, collectively called environment. This interaction generates correlations between system and environment, causing an irreversible loss of information from the system. Since there is no system that can be regarded as truly isolated, the study of the behavior of the correlations (inside the system) under the action of decoherence channels is of major importance both from theoretical and practical issues.

The usual description of the dynamics of open system is given by the so called master equation, a first order differential equation for the reduced density operator describing the system, derived by means of second order perturbation theory \cite{Breuer}. This equation, which must have a well defined Lindblad structure \cite{Lindblad}, relies on two main approximations: Weak coupling (between system and environment) and short correlation times (compared with the typical decoherence time of the system). When such approximations are not satisfied, the dynamical description of the system must change accordingly, and non-Markovian effects can appear. In classical mechanics, non-Markovian effects are identified with memory. However, in quantum mechanics, this definition is not so direct and care must be taken. We do not intend to discuss such issue here and we refer the reader to reference \cite{Non-Mark} for a recent review. In what follows we consider some results concerning the Markovian and the non-Markovian dynamical evolutions of correlations, in the context of the SCP.

\subsection{Markovian dynamics of correlations}

There is a vast literature on the subject of quantum correlations under Markovian dynamics. Here we just comment on a few results, referring the reader to more technical, and complete, references.

Shortly after the publication of Ref. \cite{Maziero_scp}, the related phenomena of quantum and classical decoherence was predicted \cite{Mazzola}. Considering energy-conserving dissipative maps (the ones described in Sec. \ref{discovery}), a class of states were identified such that before the sudden change point only the classical correlations are affected by decoherence and, after this critical point, only the quantum correlations are affected. These two regimes were then called classical and quantum decoherence \cite{Mazzola}. An elegant geometric interpretation of the sudden change behavior, for the simple case of Bell-diagonal states, was provided in Ref. \cite{Lang}, while the conditions for the correlations to stay constant, the so called freezing phenomena, were put forward in Ref. \cite{you}, considering the phase damping channel.

Following these lines, in Ref. \cite{Aaronson} it was proved that virtually all \textit{bona fide} measures of quantum correlations present the freezing effect under the same dynamical conditions.  The authors considered the case of Bell-diagonal states under the action of non-dissipative environments. A geometric interpretation was also provided.

Another interesting study was reported in Ref. \cite{Chanda}. Considering quantum correlations measured by quantum discord and quantum work deficit, the multipartite case, under local noise, was addressed. Among the results of the paper, a complementarity relation between the freezing time and the value of quantum correlation was provided. Moreover, a freezing index was introduced and its usefulness as a witness for quantum phase transition was discussed. 

Instead of correlations shared by distinct parts of the system, in Ref. \cite{MazieroSR} the authors addressed the dynamical evolution of the correlations between the system and the environment. Specifically, they considered a two-qubit system under the action of two independent channels (the Pauli and the amplitude damping maps). They found that decoherence may occur without entanglement between system and environment and also that the initial non-classical correlations, presented in the system, completely disappears, under certain conditions, being not transferred to the environments \cite{MazieroSR}.

A new interpretation of the SCP was introduced employing the idea of complementary correlations \cite{Deb}. Considering two complementary observables acting on each one of the subsystems of a bipartite system, it was shown that the sum of the local correlations between such observables is a good measure of the quantum correlations shared by the composite system. The general conclusion is that the mixedness of the initial bipartite state is not enough for the SCP, but the state also needs to present certain asymmetry with respect to local complementary observables \cite{Deb}. Moreover, they also proved that a pure state will never present the SCP and that the freezing phenomena is not a general property of all the Bell-diagonal states. It is important to mention that these results were obtained using the quantum discord as a measure for quantum correlations.

An interesting connection between the SCP and quantum teleportation was reported in Ref. \cite{Roszak}. The transition point between the classical and quantum decoherence was associated with a transition point appearing in the fidelity of the teleported state, signaling a change in the class of states that are harder to teleport. 

\subsection{Non-Markovian dynamics of correlations}

Due to memory effects, non-Markovian evolution may presents a much richer dynamics than Markovian ones. In this subsection we briefly describe some results in this field regarding the SCP.

The case of two qubits interacting with two independent non-Markovian environments was considered in Ref. \cite{Wang}, where the dynamics of entanglement was compared with that one for quantum discord. The authors verified that while entanglement can present a sudden death (entanglement disappears for all times after the critical one), quantum discord can only vanish at some specific times. In Ref. \cite{Fanchini} the authors addressed a similar problem, but now they studied the case of a common reservoir, and the SCP was again observed. An important result of this last work is the indication that the SCP is a characteristic feature of the evolution, for general initial conditions.

Regarding the freezing phenomenon, in Ref. \cite{Mazzola1} the case of two qubits under the action of local colored-noise dephasing channels was considered. As a main result, it was observed that, depending on the geometry of the initial state, the freezing phenomenon and the appearance of multiples SC was observed. Similar results were experimentally observed in Ref. \cite{Exp3}.

In Ref. \cite{ZXu} the non-Markovian dynamics of two geometric measures of quantum correlations were compared with that one for the quantum discord considering the class of Bell-diagonal states. Although all the three considered measures share a common sudden change point, one of the geometric measures does not present the freezing phenomenon. This scenario was then extended to include the treatment of the correlations between the system and the environment in Ref. \cite{Chuan-Fen}, where the appearance of the SC was studied as function of the system-environment coupling.

Relying on trace distance, several measures of quantum correlations were defined in \cite{Aaronson1}. The main result of this paper is the observation that the freezing behavior (and thus, the SCP) occurs for a larger class of states under non-Markovian dynamics than the Bell-diagonal states. 

The geometric quantum discord was employed in the development of a witness for the SCP for both Markovian and non-Markovian dynamics in Ref. \cite{CSYu}. The dependence of the freezing effect on the choice of the correlation measure were also analyzed in this work.

Reference \cite{Cianciaruso} proved that all geometric based measures of quantum correlations under the action of independent quantum channels exhibit the freezing phenomenon, and thus also the SC, for the Bell-diagonal class of two qubit states.

An interesting paper reported a study of several measures of quantum correlations, including entanglement, steering and the generalized discord, i.e. a definition of quantum correlations just like quantum discord, but based on the Tsallis $q$-entropy \cite{Renato}. In this work the authors discovered a hierarchy among all the  quantum resources and were able to identify a chronology of deaths and births (sudden changes) under non-Markovian channels. 

All of the above studies considered the case in which the environments are all identical. In Ref. \cite{BCRen} the problem of a two qubit system (considering Bell-diagonal states) under the action of distinct environments were addressed, considering both Markovian and non-Markovian dynamics. The rules governing the time evolution of the classical and quantum correlations, including the sudden change points, were established.

Studying the information flow for qubits under an Ohmic environment, in Ref. \cite{Haikka1} it was discovered a class of initial states for which quantum discord is forever frozen and the time-invariant discord was linked with non-Markovianity.

\section{The double SCP of trace distance discord}
\label{doubleSCP}

\begin{figure}[h]
\begin{centering}
\includegraphics[scale=0.5]{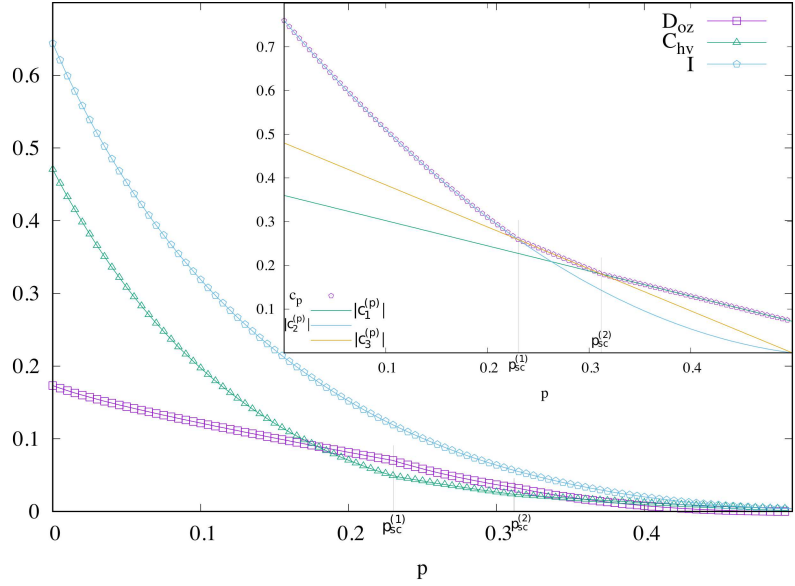} 
\par\end{centering}
\caption{Double sudden changes of Henderson-Vedral classical correlation and Ollivier-Zurek quantum discord for two qubits prepared in a Bell-diagonal state with $\mathbf{\Xi}_{0}=(0.36,-0.76,0.48)$ and with the qubits $a$ and $b$ undergoing, respectively, bit and phase flip channels. In the inset are shown the evolved components of the correlation vector and the maximal one. Here we used $r=0.8$ and the sudden change times are seen to be $p_{sc}^{(1)}\approx0.23$ and $p_{sc}^{(2)}\approx0.31$, as marked by the gray lines.}
\label{2sc_bfpf} 
\end{figure}

In this section we discuss the double sudden change phenomenon of trace distance discord ($D_{tr}$), a puzzling effect with two sudden changes (SCs) for $D_{tr}$ and only one SC for the associated classical correlation $C_{tr}$ \cite{Sarandy_2sc,Sarandy_2sc_exp}. But before doing that, let us mention that the possible existence of multiple sudden changes points for classical and quantum correlation is known since the first studies regarding their dynamics under non-Markovian global environments (see e.g. Ref. \cite{Fanchini_NM}). Also, as noticed in Ref. \cite{Deng_2sc}, two sudden change times can be obtained even with a Bell-diagonal state subject to local Markovian environments, if the qubits are acted on by suitable nonidentical surroundings. As an illustrative example, let us consider the dynamics of two qubits with the ``qubit $a$'' and ``qubit $b$'' let to evolve under the action of the bit flip (with parametrized time $p$) and phase flip (with parametrized time $q\coloneqq rp$) channels, respectively (see Sec. \ref{discovery}). Under these conditions the evolved Bell-diagonal state is determined by the correlation vector: 
\begin{equation}
\mathbf{\Xi}_{p,q}^{bf\otimes pf}=\left((1-2q)c_{1},(1-2q)(1-2p)c_{2},(1-2p)c_{3}\right).
\end{equation}
Let's assume also that the probability for the bit flip error is not greater than that for the phase flip: $q<p\therefore r\in(0,1)$. We shall have thus that the second component of $\mathbf{\Xi}_{p,q}^{bf\otimes pf}$ decays faster than the third one, which by its turn decays faster than the first one. Therefore, in this setting, any initial state with 
\begin{equation}
|c_{2}|>|c_{3}|>|c_{1}|
\end{equation}
will lead to two sudden changes for $C_{hv}$ and $D_{oz}$ if the crossing $|c_{2}^{(p,q)}|=|c_{3}^{(p,q)}|$ happens before than the intersection $|c_{2}^{(p,q)}|=|c_{1}^{(p,q)}|$ does. This leads to the following additional requirement for the existence of a double sudden change: 
\begin{equation}
r>\frac{|c_{2}|-|c_{3}|}{|c_{2}|-|c_{1}|}.
\end{equation}
Once all these conditions are fulfilled and $p\in[0,2^{-1}]$ is adopted, we will see the SCP taking place at the times 
\begin{equation}
p_{sc}^{(1)}=\frac{|c_{2}|-|c_{3}|}{2r|c_{2}|}\mbox{ and }p_{sc}^{(2)}=\frac{|c_{3}|-|c_{1}|}{2(|c_{3}|-r|c_{1}|)},
\end{equation}
as exemplified in Fig. \ref{2sc_bfpf}. We remark that similar conditions can be obtained in an analogous manner for other combinations of local channels. Besides, we observe that if $r=1\therefore q=p$ then $p_{sc}^{(2)}=2^{-1}$ and there is only one SCP.

Now we shall address the double sudden change phenomenon (DSCP) \emph{per se}. In order to do that we will make use of the analytical formulas \cite{Giovannetti_AnalDtr,Sarandy_epb} for the trace distance correlations of two-qubit X states: 
\begin{equation}
\rho^{x}=\begin{bmatrix}\rho_{11} & 0 & 0 & |\rho_{14}|\mathrm{e}^{i\phi_{14}}\\
0 & \rho_{22} & |\rho_{23}|\mathrm{e}^{i\phi_{23}} & 0\\
0 & |\rho_{23}|\mathrm{e}^{-i\phi_{23}} & \rho_{33} & 0\\
|\rho_{14}|\mathrm{e}^{-i\phi_{14}} & 0 & 0 & \rho_{44}
\end{bmatrix}.
\end{equation}
This class of state can be transformed into the standard form 
\begin{equation}
\rho^{xr}=\begin{bmatrix}\rho_{11} & 0 & 0 & |\rho_{14}|\\
0 & \rho_{22} & |\rho_{23}| & 0\\
0 & |\rho_{23}| & \rho_{33} & 0\\
|\rho_{14}| & 0 & 0 & \rho_{44}
\end{bmatrix}
\end{equation}
by applying the following local unitary (LU) transformation: 
\begin{equation}
U=U_{a}\otimes U_{b}=\mathrm{e}^{-i(\phi_{14}+\phi_{23})\sigma_{3}/4}\otimes\mathrm{e}^{-i(\phi_{14}-\phi_{23})\sigma_{3}/4}.
\end{equation}
As a basic property required for correlation quantifiers is invariance under local unitaries, we can study the correlations of $\rho^{x}$ using its LU-equivalent version $\rho^{xr}$. When dealing with distance-based correlation measures, it is frequently useful utilizing the Bloch representation for the density operator; which in this case is 
\begin{equation}
\rho^{xr}=2^{-2}\left(\mathbb{I}^{ab}+a_{3}\sigma_{3}\otimes\mathbb{I}^{b}+b_{3}\mathbb{I}^{a}\otimes\sigma_{3}+{\textstyle \sum_{j=1}^{3}}c_{jj}\sigma_{j}\otimes\sigma_{j}\right),
\end{equation}
where the non-null elements of the Bloch vectors and correlation matrix are 
\begin{eqnarray}
a_{3} & = & \mathrm{Tr}(\rho^{xr}\sigma_{3}\otimes\mathbb{I}^{b})=2(\rho_{11}+\rho_{22})-1,\\
b_{3} & = & \mathrm{Tr}(\rho^{xr}\mathbb{I}^{a}\otimes\sigma_{3})=2(\rho_{11}+\rho_{33})-1,\\
c_{11} & = & \mathrm{Tr}(\rho^{xr}\sigma_{1}\otimes\sigma_{1})=2(|\rho_{23}|+|\rho_{14}|),\label{eq:c11_rhox}\\
c_{22} & = & \mathrm{Tr}(\rho^{xr}\sigma_{2}\otimes\sigma_{2})=2(|\rho_{23}|-|\rho_{14}|),\\
c_{33} & = & \mathrm{Tr}(\rho^{xr}\sigma_{3}\otimes\sigma_{3})=1-2(\rho_{22}+\rho_{33}).
\end{eqnarray}
We observe that the effect of the LU above when transforming $\rho^{x}$ into $\rho^{xr}$ is: $c_{11}:2(\mathrm{Re}\rho_{23}+\mathrm{Re}\rho_{14})\rightarrow2(|\rho_{23}|+|\rho_{14}|)$, $c_{22}:2(\mathrm{Im}\rho_{23}-\mathrm{Im}\rho_{14})\rightarrow0$, $c_{22}:2(\mathrm{Re}\rho_{23}-\mathrm{Re}\rho_{14})\rightarrow2(|\rho_{23}|-|\rho_{14}|)$, and $c_{21}:-2(\mathrm{Im}\rho_{23}+\mathrm{Im}\rho_{14})\rightarrow0.$

Regarding the analytical formulas for the correlations, the trace distance discord of X states was obtained in Ref. \cite{Giovannetti_AnalDtr} and reads 
\begin{eqnarray}
D_{tr}^{a}(\rho^{x}) & \coloneqq & \min_{\rho_{cq}}d_{tr}(\rho,\rho_{cq})\\
 & = & \sqrt{\frac{\max\left(c_{33}^{2},a_{3}^{2}+\min\left(c_{11}^{2},c_{22}^{2}\right)\right)\max\left(c_{11}^{2},c_{22}^{2}\right)-\min\left(c_{33}^{2},\max\left(c_{11}^{2},c_{22}^{2}\right)\right)\min\left(c_{11}^{2},c_{22}^{2}\right)}{\max\left(c_{33}^{2},a_{3}^{2}+\min\left(c_{11}^{2},c_{22}^{2}\right)\right)-\min\left(c_{33}^{2},\max\left(c_{11}^{2},c_{22}^{2}\right)\right)+\max\left(c_{11}^{2},c_{22}^{2}\right)-\min\left(c_{11}^{2},c_{22}^{2}\right)}}.
\end{eqnarray}
Closed formulas were obtained in Ref. \cite{Sarandy_epb} for the trace distance classical and total correlations: 
\begin{eqnarray}
C_{tr}^{a}(\rho^{x}) & \coloneqq & \max_{\Pi_{a}}d_{tr}(\Pi_{a}(\rho^{x}),\Pi_{a}(\rho_{a}^{x}\otimes\rho_{b}^{x}))=\kappa_{+},\\
I_{tr}(\rho^{x}) & \coloneqq & d_{tr}(\rho^{x},\rho_{a}^{x}\otimes\rho_{b}^{x})=\frac{1}{4}\sum_{j,k=0}^{1}|c_{11}+(-1)^{j}c_{22}+(-1)^{k}(c_{33}-a_{3}b_{3})|=\frac{1}{2}\left(\kappa_{+}+\max(\kappa_{+},\kappa_{0}+\kappa_{-})\right).
\end{eqnarray}
Above, $\kappa_{+}$, $\kappa_{0}$ and $\kappa_{-}$ are, respectively, the maximum, intermediate, and minimum value among the elements in the list \{$|c_{11}|,|c_{22}|,|c_{33}-a_{3}b_{3}|\}$, $\Pi_{a}$ is a complete non-selective measurement map, and $\rho_{a(b)}^{x}=\mathrm{Tr}_{b(a)}(\rho^{x})$ are the reduced density matrices. Its is interesting noticing here that for X states this definition of trace distance classical correlation is symmetric, $C_{tr}^{a}(\rho^{x})=C_{tr}^{b}(\rho^{x})$, while the associated discord measure is generally asymmetric \cite{Maziero_DiscAn}, $D_{tr}^{a}(\rho^{x})\ne D_{tr}^{b}(\rho^{x})$, if $a_{3}\ne b_{3}$.

\begin{figure}[h]
\begin{centering}
\includegraphics[scale=0.5]{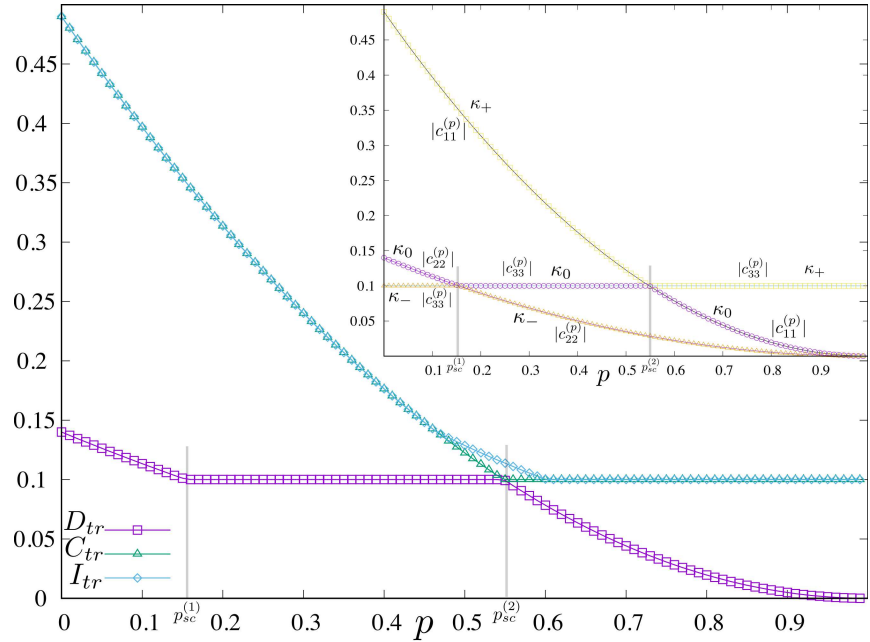} 
\par\end{centering}
\caption{Double sudden change phenomenon of trace distance quantum discord for Bell-diagonal states (with $(c_{11},c_{22},c_{33})=(0.49,-0.14,-0.10)$) evolving under the action of local independent phase damping channels. As shown by the gray lines, there are two sudden changes of $D_{tr}$, at $p_{sc}^{(1)}\approx0.16$ and at $p_{sc}^{(2)}\approx0.55$, while the decaying rate of the classical correlation $C_{tr}$ changes abruptly only at $p_{sc}^{(2)}$. It is noteworthy that the total correlation is (nearly) equal to the classical correlation for most values of $p$; a fact that may motivate the questioning about the suitability of these correlations definitions. Moreover, we identify the existence of abrupt changes also for $I_{tr}$ at $p_{\pm}=1-\sqrt{|c_{33}|/(|c_{11}|\pm|c_{22}|)}$ (for this initial state $p_{-}\approx0.47$ and $p_{+}\approx0.60$). As the definition of $I_{tr}$ entails not extremizations at all, this effect has to be induced by the trace distance itself.}
\label{2sc_tr} 
\end{figure}

Next we consider the dynamics of $\rho^{xr}$ under local phase damping channels, which maintain the X form for $\rho^{x}$, and for its LU equivalent $\rho_{p}^{xr}$, with the only changes: 
\begin{equation}
c_{11}\rightarrow c_{11}(1-p)^{2}\mbox{ and }c_{22}\rightarrow c_{22}(1-p)^{2};
\end{equation}
all the other parameters remain constant with time. For easily seeing the mathematical origin of the DSCP, let us look at the special case of Bell-diagonal states ($a_{3}=b_{3}=0$). For these states $D_{tr}(\rho_{p}^{bd})$ and $C_{tr}(\rho_{p}^{bd})$ are given, respectively, by the intermediate and maximum values among those in the list 
\begin{equation}
(|c_{11}^{(p)}|,|c_{22}^{(p)}|,|c_{33}^{(p)}|)=(|c_{11}|(1-p)^{2},|c_{22}|(1-p)^{2},|c_{33}|).
\end{equation}
Thus, the analysis of the sudden changes of $C_{tr}$ is equivalent to that for $C_{hv}$ (see Sec. \ref{discovery}). On the other side, we readily see that for a Bell-diagonal initial state with 
\begin{equation}
|c_{11}|>|c_{22}|>|c_{33}|
\end{equation}
at the times 
\begin{equation}
p_{sc}^{(1)}=1-\sqrt{\frac{|c_{33}|}{|c_{22}|}}\mbox{ and }p_{sc}^{(2)}=1-\sqrt{\frac{|c_{33}|}{|c_{11}|}}
\end{equation}
the intermediate value changes, and so does the decay rate of $D_{tr}$. In summary, the number of SCs of $D_{tr}$ is greater than the number of abrupt changes of $C_{tr}$. This is the main feature of the DSCP, which is exemplified in Fig. \ref{2sc_tr}. We notice that the analysis for $X$ states can be made in a similar fashion, just by replacing $c_{33}$ with $c_{33}-a_{3}b_{3}$.

\begin{figure}[h]
\begin{centering}
\includegraphics[scale=0.4]{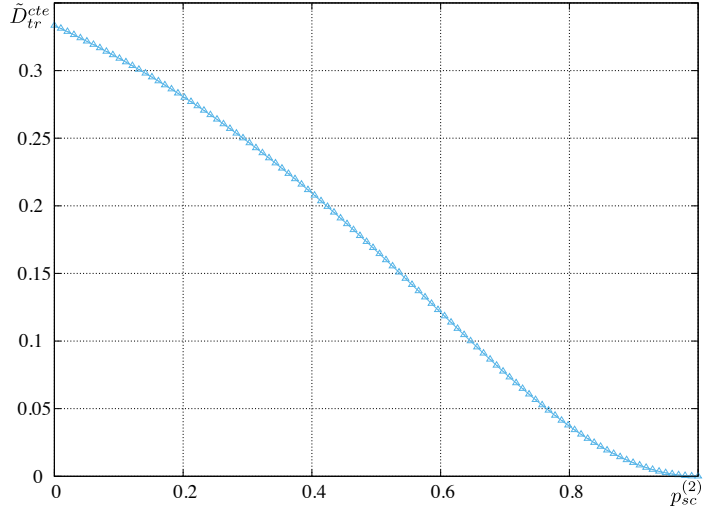} 
\par\end{centering}
\caption{Upper bound for the constant value of trace distance discord as a function of the second sudden change time. This bound holds for any Bell-diagonal state with $|c_{22}|=|c_{33}|$ and evolving under local-independent phase damping channels.}
\label{fig:ub_tdd} 
\end{figure}

As can be seen in Fig. \ref{2sc_tr}, $D_{tr}(\rho_{p}^{bd})=|c_{33}|$ is constant, or freezed, for $p\in(p_{sc}^{(1)},p_{sc}^{(2)})$. Given the recognized utility of quantum discord as a resource for some tasks in quantum information science, its robustness to noise is a most welcome feature; one that we should try to maintain for how much time as we can. In the case of Bell-diagonal states described above, we can begin doing that by choosing $|c_{22}|=|c_{33}|$, what implies in $p_{sc}^{(1)}=0$. The next step is trying to get $p_{sc}^{(2)}$ as close to one as possible. But we perceive that the positivity of $\rho^{bd}$ restricts the possible values of $c_{33}$ and $c_{11}$; and this will lead to the following tradeoff between $p_{sc}^{(2)}$ and the constant value of $D_{tr}$: 
\[
D_{tr}^{cte}\le\frac{(1-p_{sc}^{(2)})^{2}}{1+2(1-p_{sc}^{(2)})^{2}}\eqqcolon\tilde{D}_{tr}^{cte}.
\]
Therefore, for instance, $p_{sc}^{(2)}=2^{0}\Rightarrow D_{tr}^{cte}=0$, $p_{sc}^{(2)}=2^{-1}\Rightarrow D_{tr}^{cte}\le0.16$, and $p_{sc}^{(2)}=2^{-2}\Rightarrow D_{tr}^{cte}\le0.26$. The upper bound $\tilde{D}_{tr}^{cte}$ for the constant value of TDD is shown in Fig. \ref{fig:ub_tdd} for $p_{sc}^{(2)}\in[0,1]$.

The double sudden change phenomenon, which has been considered to be a genuine quantum feature \cite{Sarandy_2sc_exp}, is a very interesting effect. Notwithstanding, as discussed in this section, there are several open questions motivated by the analysis of this phenomenon. Some examples of such questions are: Can we give a physical interpretation to the DSCP? Isn't the DSCP simply a byproduct of the distance measure utilized? Shouldn't we further analyze which properties a distance measure should have to be used in correlations quantification \cite{Maziero_NMuTP}? For which other distance or distinguishability measures this kind of effect can be observed \cite{Wei_2sc,Orszag_2sc}? We think that in addressing these questions, we can deep our understanding about quantum systems and its correlations.

\section{Final remarks, open questions, and perspectives}
\label{conclusions}

The sudden change phenomenon of quantum discord is an interesting and elusive effect. Despite the considerable amount of work dedicated or related to it have been constructing some knowledge, we are yet faced with more open questions than answers. Some of the main issues we leave here are:

Symmetry is a central concept in physics and is central e.g. for conservations laws. Can the resource theory of asymmetry bring to light the possible relevance of the SCP in physics?

Quantum discord has been shown to be a surprisingly general  ``order parameter'' for phase transitions. As the sudden modification of its rate of change with the physical parameter determining the system state (phase) seems to play a key role here, it would be fruitful to pursue more formal connections between the SCP and symmetry (breaking) in phase transitions.

It is now clear that composite-discordant quantum states are the resource making possible the efficient implementation of several tasks in quantum information science. And, in our ever going fight against some noise interactions, the complete robustness of QD to environment influences is a most welcome feature it possesses. On the other side, one can also identify the existence of tradeoffs between the constant value of QD and the amount of time it can stay unchanged. It is important thus to analyze what kind of limitation this last effect can have on the possible practical applications of quantum discord. 

Also, from the overall analysis we made here, on one side we can conclude that, as different discord functions are related to different physical or operational quantities, then the interpretation of the SCP could also be measure dependent. On the other hand, if future investigations happen to give a universal physical meaning to the SCP, then this fact could be considered in order to infer further conditions discord functions should have to satisfy to be faithfully used for the sake of quantum correlations quantification.

\begin{acknowledgments}
We acknowledge financial support from the Brazilian funding agencies CNPq (Grants No. 401230/2014-7, 445516/2014-3, 305086/2013-8, 441875/2014-9 and 303496/2014-2) and CAPES (Grant No. 6531/2014-08), the Brazilian National Institute of Science and Technology of Quantum Information (INCT/IQ). JM gratefully acknowledges the hospitality of the Physics Institute and Laser Spectroscopy Group at the Universidad de la Rep\'{u}blica, Uruguay. 
\end{acknowledgments}



\end{document}